\documentclass[prb,twocolumn,aps,showpacs,superscriptaddress]{revtex4}
\usepackage{color}
\usepackage{graphicx}
\def\be{\begin{equation}}
\def\ee{\end{equation}}

\begin{document}

\title{Logarithmic scaling of Lyapunov exponents in disordered chiral two-dimensional lattices}

\author{P. Marko\v{s}}
\affiliation{Dept.\ Physics, FEI\,STU, 812\,19 Bratislava, Slovakia} 
\author{ L. Schweitzer}
\affiliation{Physikalisch-Technische Bundesanstalt (PTB), Bundesallee 100,
  38116  Braunschweig, Germany} 

\begin{abstract}
We analyze the scaling behavior of the two smallest Lyapunov exponents for
electrons propagating on two-dimensional lattices with energies within a
very narrow interval around the chiral critical point at $E=0$ in the presence
of a perpendicular random magnetic flux.  
By a numerical analysis of the energy and size dependence we confirm 
that the two smallest Lyapunov exponents are functions of
a single parameter. The latter is given by $\ln L/\ln \xi(E)$, which is
the ratio of the logarithm of the system width $L$ to the logarithm of the
correlation length $\xi(E)$. Close to the chiral critical point and energy 
$|E|\ll E_0$, we find a logarithmically divergent energy dependence
$\ln \xi(E)\propto |\ln(E_0/|E|)|^{1/2}$, where $E_0$ is a characteristic
energy scale.  
Our data are in agreement with the theoretical prediction of M.\ Fabrizio and
C.\ Castelliani [Nucl.\ Phys.\ B {\bf 583},  542 (2000)] and resolve an
inconsistency of previous numerical work. 
\end{abstract}
\pacs{05.30.Rt, 61.43.Bn, 71.23.An, 71.55.Jv}
\maketitle

\section{Introduction}
The numerical determination of transport parameters for electrons propagating
in disordered two-dimensional systems with chiral symmetry still remains an
important unsolved problem. The situation can be represented by a single-band
tight-binding model defined on bipartite lattices subjected to purely
off-diagonal disorder like a random-magnetic flux with zero mean or real
random hopping terms. The latter belongs to the chiral orthogonal universality
class while the former is chiral unitary.
Due to the chiral symmetry, the model exhibits metallic behavior only at energy
$E=0$.\cite{EM08} It is therefore of considerable interest to study the critical
properties of the model in the vicinity of the critical point and to
investigate its universality.

The traditional finite-size-scaling analysis of disorder driven metal-insulator
transitions, i.e., continuous quantum phase transitions at zero temperature,
is based on two assumptions.\cite{AALR79,PS81,PS81a,MK81} (i) In the vicinity
of the critical point all variables of interest are a function of only one
parameter, 
\be\label{B}
v(E,L) = F(L/\xi(E)),
\ee
where $\xi(E)$ is the energy dependent correlation length. Here, vicinity
means that both the system size $L$ and $\xi(E)$ are already larger than any
other typical length of the model and $\xi>L$. 
(ii) At the critical energy $E_c$, the correlation length $\xi(E)$ diverges as 
\be\label{A}
\xi(E)\sim |E-E_c|^{-\nu}
\ee
with a universal critical exponent $\nu$.
Relation (\ref{A}) was confirmed in a multitude of numerical work on
disordered systems in spatial dimension $2\le d\le 5$ and various
physical symmetries.\cite{Mar06} As scaling variables, for example,
the localization length,\cite{Mac80} the smallest Lyapunov
exponent,\cite{SO99} the two-terminal conductance,\cite{SMO01,MS06} the energy
level spacings,\cite{Sea93,SZ97,PS02} and the inverse participation
ratio\cite{BM06} were successfully used.   

Recently, the analysis of numerical data obtained
for the energy dependence of the two-terminal conductance $g$ on a bricklayer
lattice,\cite{SM08a} which represents a generic lattice model for graphene, has  
led to a power-law energy dependence of the correlation length $\xi(E) 
\propto |E|^{-\nu}$, where the critical exponent $\nu$ is close to $1/3$.   
Although this outcome is in agreement with previous numerical
results\cite{Cer00,ERS01,Cer01,ER04,MS07} for square and hexagonal lattices, 
it is at variance with the Harris criterion \cite{Har74} which states that
$\nu>1/d$, where $d=2$ is the euclidian dimension of the system. More
importantly, all numerical data obtained to the present date do not agree 
with theoretical predictions,\cite{FC00,EM08} according to which the
correlation length depends logarithmically on the energy,  
\begin{equation}\label{fc}
\xi(E) = \xi_0 \exp[A\sqrt{\ln(E_0/|E|)}],\quad |E|\ll E_0,
\end{equation}
where $A$ is related to the longitudinal conductivity and $E_0$ is assumed to
be of the order of the energy band width.\cite{EM08}
A possible explanation of this disagreement between theory and numerical
experiments may be that the energies investigated in the numerical
studies, down to $10^{-10}$ so far\cite{MS07} (in units of the hopping energy),
are not sufficiently small in comparison to the unspecified 
parameter $E_0$ introduced in the theory. Thus, it could be that the energy
interval $|E|\ll E_0$, where the scaling holds, was not reached in previous
numerical studies. A second obstacle is the vanishing of the density of states
$\rho(E)$ which occurs at $E=0$ for hexagonal and bricklayer lattices in the
presence of random-magnetic-flux disorder.\cite{Sch09} This behavior persists
even in strongly disordered chiral systems so that the two-terminal
conductance, which nevertheless turns out to be finite $\sim e^2/h$ at the
Dirac-point in graphene,\cite{LFSG94,Zie07,OGM07,SM08a} is not a suitable
scaling variable for numerical studies. Therefore, it is expedient to
investigate instead the smallest Lyapunov exponents, which are associated with the
localization length and are not directly affected by the vanishing density of
states.   

In this paper, we analyze the scaling behavior of both the two-dimensional
bricklayer and square-lattice model with random-magnetic-flux disorder. Using
the transfer-matrix method for quasi-1d systems,\cite{PS81,PS81a,MK81,SO99} 
we calculate the two smallest Lyapunov exponents $z_1$ and $z_2$ for energies
very close to $E=0$, not achieved in previous work. 
Lyapunov exponents are more suitable than the conductance since they are less
sensitive to the energy dependence of the density of states. Another reason
for using Lyapunov exponents is that the analysis of the conductance is far
more time consuming, which is crucial since quadruple precision is 
necessary in our case when energies smaller than $10^{-16}$ are considered. 
We show that in the vicinity of the critical point our numerical data for the
Lyapunov exponents lead to the relation  
\be\label{scal-1}
z_{1,2}(E,L) = F_{1,2}\left(\frac{\ln L}{\ln\xi(E)}\right)
\ee
where $L$ is the width of the system. We prove that the correlation length
$\xi(E)$ depends logarithmically on the energy (see Eq.~(\ref{fc}))
in agreement with the predictions by Fabrizio and Castelliani.\cite{FC00}
Thus, our results resolve the previous discrepancy between analytical theory
and numerical calculations.

\section{The model and method}
We study a single-band tight-binding Hamiltonian defined on a two-dimensional
square lattice with nearest-neighbor hopping and random-flux disorder, which
is introduced by complex phase factors in the transfer terms,
\begin{eqnarray}
{\cal H}/V &=& \sum_{x,y}{}^{'} \big(e^{i\theta_{x,y+a;x,y}} c_{x,y}^\dag
c_{x,y+a}\nonumber \\
&&
+ e^{-i\theta_{x,y-a;x,y}} c_{x,y}^\dag c_{x,y-a}\big)
\nonumber \\
& &
+ \sum_{x,y} \big(c_{x,y}^\dag c_{x+a,y}+  c_{x,y}^\dag c_{x-a,y}\big),
\label{ham}
\end{eqnarray}
where $c_{x,y}^\dag$ and $c_{x,y}$ denote creation and annihilation operators
of a fermionic particle at site $(x,y)$, respectively. For bricklayer lattices,
the prime at the first sum in (\ref{ham}) indicates that only the transfers
along every other vertical bond are included. In this way, the square lattice
is transformed into a bricklayer where the coordination number is reduced to
three nearest-neighbor sites. The bricklayer lattice has the same topology as
the honeycomb lattice of graphene and the Hamiltonian (\ref{ham}) possesses
the same eigenvalues $\pm \varepsilon_i$.  The phases, which are chosen to be  
associated only with the vertical bonds in the $y$-direction,
$\theta_{x,y;x,y+a}=\theta_{x+2a,y;x+2a,y+a}-\frac{2\pi e}{h}\phi_{x,y}$,
are defined by the random magnetic flux $\phi_{x,y}$,  which is uniformly
distributed $-f/2\le\phi_{x,y}\le f/2$ with zero mean and disorder strength
$0\le f/(h/e)\le 1$. The random magnetic flux is pointing perpendicular to
the 2d lattice and periodic boundary conditions are applied in the
$y$ direction. In contrast to diagonal disorder, this random flux
preserves the chiral symmetry for both the square and bricklayer lattices.
We fix the units of energy and length scales by the nearest-neighbor hopping
energy $V=1$ and the lattice constant $a=1$, respectively. The disorder
strength is taken to be $f=0.5~h/e$ for the bricklayer and $f=1.0~h/e$ for
the square lattice.

We use the transfer matrix method\cite{Mar06} and collect numerical data for
the two smallest Lyapunov exponents $z_1(E,L)$ and $z_2(E,L)$.  For the system
width $L$ and length $L_x \gg L$, we calculate the transfer matrix ${\bf M} =
\prod_i^{L_x} {\bf M}_i$, and extract the two smallest Lyapunov exponents. The
relative uncertainty $\epsilon(E,L)$ of our data is $2\cdot 10^{-3}$ for larger
widths $L=192$  and $L=160$, and decreases down to $10^{-4}$ for the smallest
$L=8$. This requires the length of the quasi-1d systems $L_x$ to be in the
range $\sim 10^8$ to  $10^9$. Since we expect that scaling occurs only in the
vicinity of the $E=0$ critical point, we consider energies as small as
possible, down to the point of $|E| = 10^{-34}$, at least for $L\le 64$. This
requires to perform the calculations with quadruple numerical precision.

\begin{figure}
\includegraphics[clip,width=0.43\textwidth]{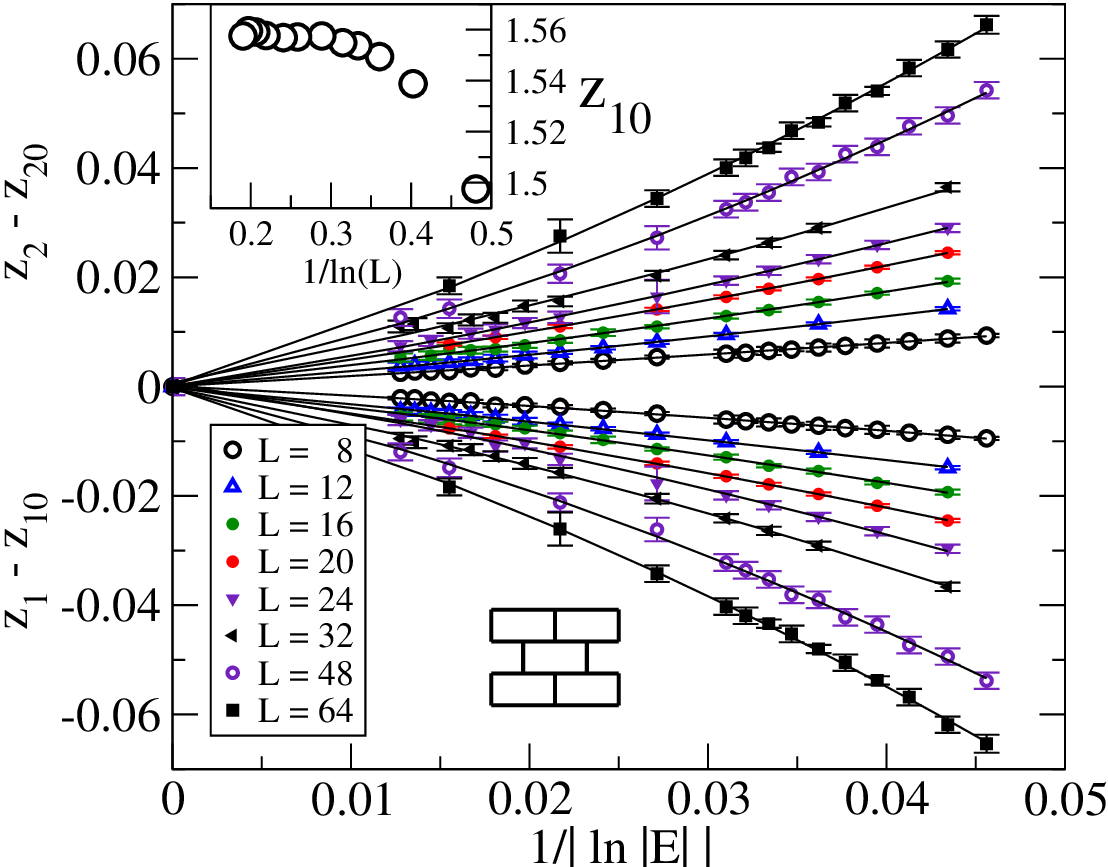}
\caption{(Color online) The smallest Lyapunov exponents $z_1(E,L)-z_1(E=0,L)$
  (lower branches) and $z_2(E,L)-z_2(E=0,L)$ (upper branches) as a function of
  $1/|\ln |E||$ for energies $|E|<3\cdot 10^{-10}$. 
  The applied random flux strength is $f=0.5 h/e$ 
  and the width of the quasi-1d systems is in the
  range $8\le L \le 64$. Solid lines are quadratic fits.
  The inset shows the size dependence of $z_{10}=z_1(E=0,L)$ for $8\le L\le 192$. 
}
\label{hx_32}
\end{figure}

The specific symmetry of the model provides us with an independent test of the
accuracy of our data. Due to the chiral symmetry, the spectrum of Lyapunov
exponents must be degenerate at the band center for all $L$, 
\be\label{e0}
z_1(E=0,L)=z_2(E=0,L).
\ee
Deviations from $E=0$ remove this degeneracy but the average value,
$[z_1(E)+z_2(E)]/2$, equals to $z_1(E=0)$ for small values of $E$. 

\begin{figure}
\includegraphics[clip,width=0.33\textwidth]{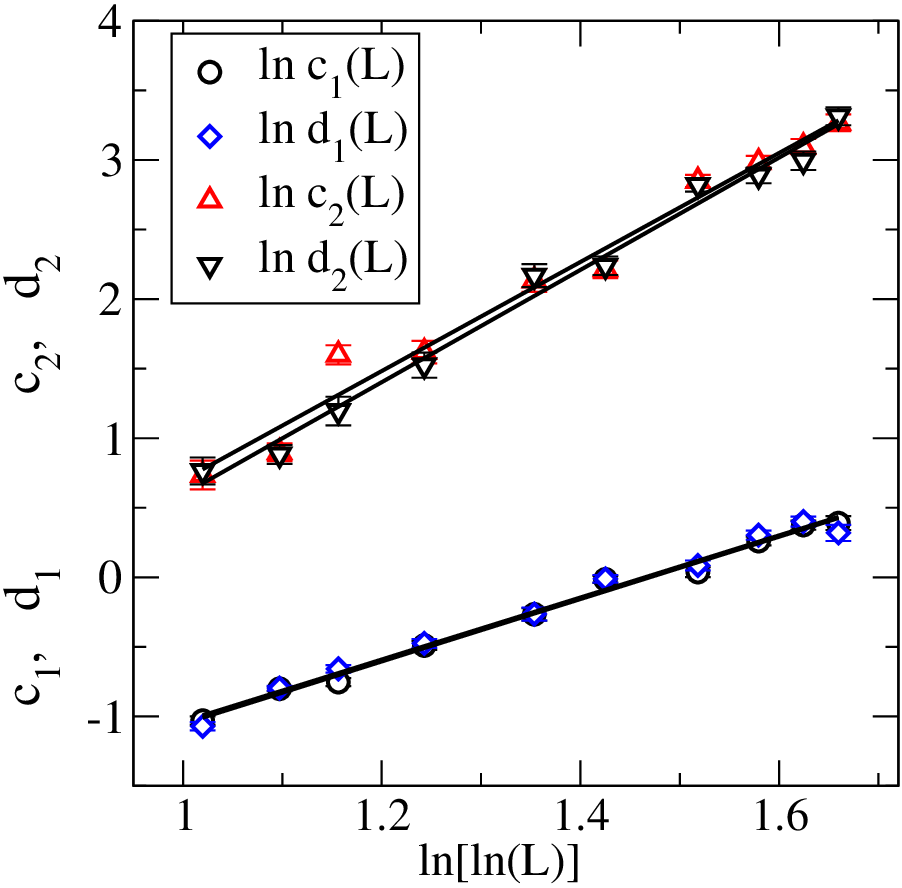}
\caption{(Color online) The size dependence of the coefficients $c_i$ and $d_i$,
  given by (\ref{c1}) and (\ref{c2}). Solid lines are linear fits with slopes
  $2.24\pm  0.26$, $2.23\pm 0.26$ ($c_1$ and $d_1$), and $3.92\pm 0.38$,
  $4.03\pm 0.40$ ($c_2$ and $d_2$). The width is in the range $16\le L\le 192$
  and  $E_0=0.8$. Only data for $|E|\le 3\times 10^{-10}$ were used. 
}
\label{h5_cL_log}
\end{figure}

\section{Bricklayer lattice}
The energy dependence of the two Lyapunov exponents is plotted in
Fig.~\ref{hx_32} for various system widths $L$ of the bricklayer. Our
data confirm that $z_1$ and $z_2$ are analytical functions of the variable 
$1/|\ln |E||$. Therefore, we approximate their energy dependence by the
Taylor expansion 
\be\label{taylor}
z_{1}(\chi,L)-z_1(0,L)  = c_0(L) + c_1(L) \chi + c_2(L) \chi^2,
\ee
where 
\be\label{chi}
\chi = \frac{1}{|\ln (E_0/|E|)|},
\ee
and by a similar expansion $z_2(\chi,L) - z_2(0,L) = d_2(0,L) + d_1(L) \chi + d_2(L)
\chi^2$, for the second Lyapunov exponent. 
Comparing with Eq.~(2), we conclude that $\ln (\xi_0/a) \ll A [\ln (E_0/|E|)]^{1/2}$ 
so that in what follows we consider $\xi_0 \sim a$.
The expansion coefficients $c_i$ and $d_i$ are determined numerically. 
The $L$-dependence of the coefficient $c_0$ and $d_0$ shows finite size
corrections. For the bricklayer, we found that $c_0$ and $d_0$ depend only
weakly on $L$ provided that $L>16$ (data are shown in the inset of
Fig.~\ref{hx_32}). For instance, we obtain that $z_1(E=0,L=16) = 1.5498\pm
0.0003$ and $z_1(E=0,L=192) = 1.557\pm 0.002$.   

\begin{figure}
\includegraphics[clip,width=0.4\textwidth]{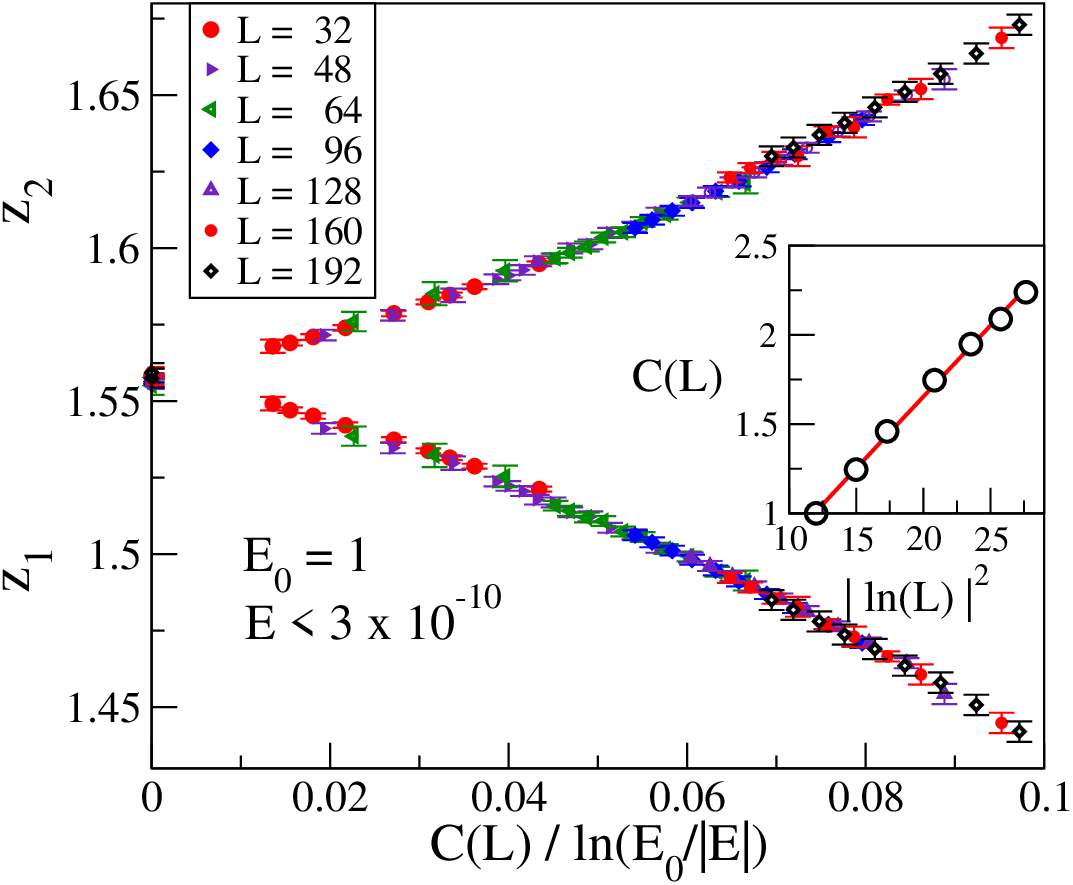}
\caption{(Color online) Finite-size scaling of the two lowest Lyapunov
  exponents for various widths $L$ and restricted energies $|E|<3\cdot
  10^{-10}$. The inset shows the $L$-dependence of the parameter $C(L)$. The 
  solid line is the power-law fit confirming that $C(L) \propto (\ln
  L)^{1.934}$. 
}
\label{hx_32a}
\end{figure}

To estimate the energy $E_0$, we first minimize the expression
\be\label{XX}
X=\sum_{E,L} \frac{[z(E,L)-F(E,L)]^2}{[z(E,L)\epsilon(E,L)]^2},
\ee
where
\be
F(E,L) = \beta_0+\beta_1(\ln L)^{\alpha_1}\chi + \beta_2(\ln L)^{\alpha_2}\chi^2,
\ee
with respect to parameters $\alpha$, $\beta$, and $E_0$. We found that $X$
possesses a minimum when $2.2<\alpha_1<2.5$, $3.8< \alpha_2<4.1$, and $0.1<
E_0$. It was not possible to obtain a better estimation of the critical
parameters from this procedure since small variation of $E_0$ can be
compensated by small change of $\beta_2$ and $\alpha_2$. 
Instead, in a more accurate analysis, we fit our numerical data for $z_1$ and
$z_2$ to the quadratic expansion (\ref{taylor}). 
Fig.~\ref{h5_cL_log}
shows the $L$-dependence of the coefficients $c_{1,2}$
and $d_{1,2}$. Our data confirm the assumed logarithmic behavior of all 
coefficients occurring in the expansion 
\be\label{c1}
c_1(L),~~~ d_1(L) \propto \left[\ln(L)\right]^{\alpha_1}
\ee
and
\be\label{c2}
c_2(L),~~~ d_2(L) \propto \left[\ln(L)\right]^{\alpha_2},
\ee
where 
$\alpha_1$ and $\alpha_2$ are close to the anticipated values $2$ and $4$,
respectively. 

Fig.~\ref{hx_32a} shows another test of the scaling of the Lyapunov
exponents. 
Following the conventional scaling method,\cite{MK81} we re-scaled the
horizontal axis for the data shown in Fig.~\ref{hx_32} by the parameter
$C(L)$: $\chi\to \chi C(L)$. The such obtained $C(L)$ gives us directly the
required scaling behavior as shown in the Fig.~\ref{hx_32a}.  The data for
both $z_1$ and $z_2$ scale to one universal curve. The inset to
Fig.~\ref{hx_32a} confirms the expected power-law relation $C(L)\propto (\ln
L)^2$.  

\begin{figure}
\includegraphics[clip,width=0.33\textwidth]{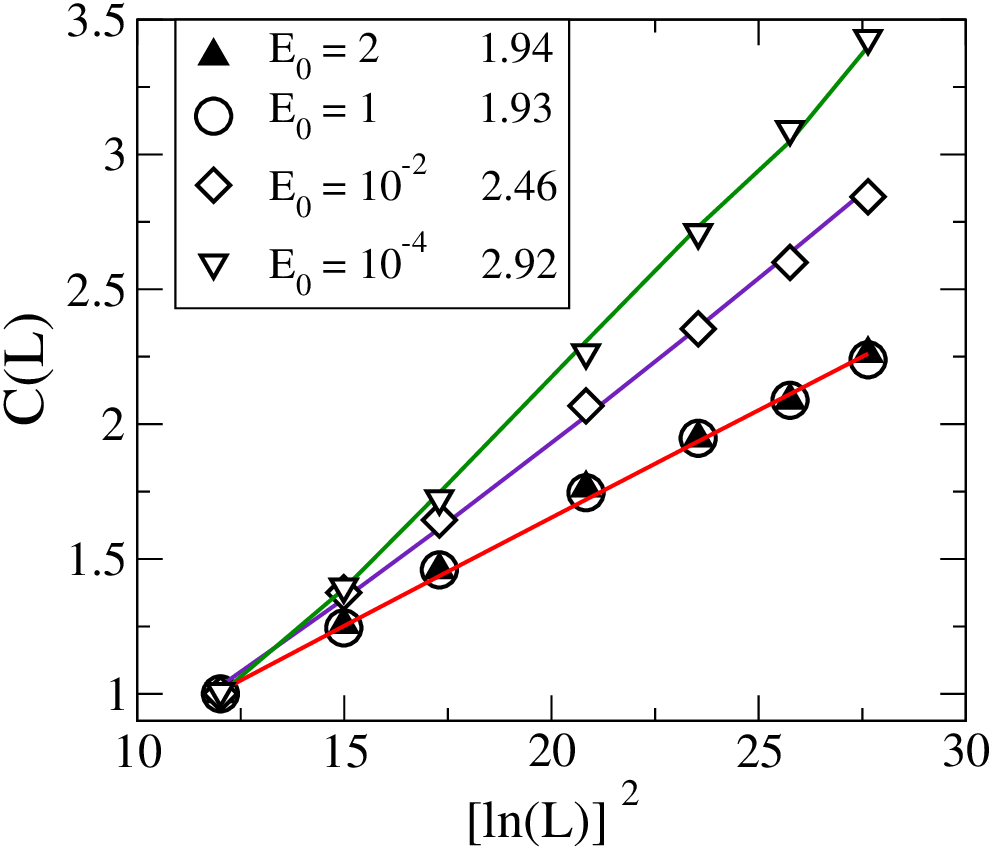}
\caption{(Color online) 
The function $C(L)$ obtained by a scaling analysis of the data for four
different values of the energy $E_0$. Solid lines are power-law fits
$C(L)\propto (\ln L)^\kappa$ with the value of the exponent $\kappa$ given in
the legend.  
}
\label{h5_cL}
\end{figure}

To obtain a quantitative estimation of the energy $E_0$, we repeated the
scaling analysis shown in Fig.~\ref{hx_32a} for  various $E_0$.  
Although we recovered the scaling behavior similar to that shown in
Fig.~\ref{hx_32} (data not shown), the $L$-dependence of the parameter $C(L)$
depends on the choice of $E_0$. As shown in Fig.~\ref{h5_cL},
$C(L) \propto (\ln L)^{\kappa}$ with the exponent $\kappa$ decreasing when
$E_0$ increases, converging to $\kappa\approx 2$ for $E_0\gtrsim 1$.
We conclude that the energy $E_0$ is of the order of unity in our bricklayer
model. 

\begin{figure}
\includegraphics[clip,width=0.4\textwidth]{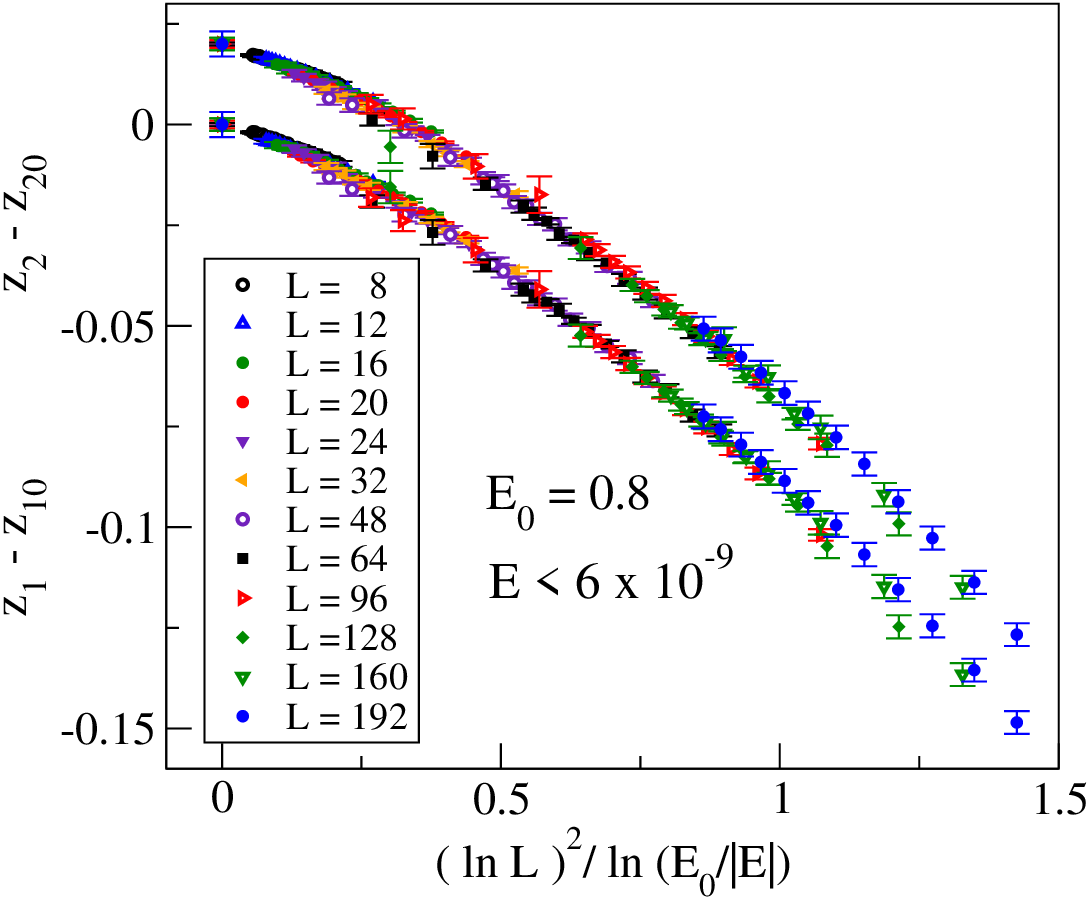}
\caption{(Color online) 
The data-collapse of the first two Lyapunov exponents $z_1(E,L)- z_{10}(L)$ 
and $z_2(E,L) - z_{20}(L)$ plotted as a function of a single parameter
$(\ln L)^2/\ln (E_0/|E|)$ with $E_0=0.8$.  
For clarity, the data for the second Lyapunov exponent are shifted vertically.
}
\label{figX}
\end{figure}

Finally, we plot in Fig.~\ref{figX} the two smallest Lyapunov exponents as 
a function of a single parameter $(\ln L)^2/\ln (E_0/|E|)$  with $E_0=0.8$. 
To reduce  the finite size corrections, we subtract from the data the 
values $z_1(E=0.L)$ and  $z_2(E=0,L)$, respectively. All data collapse onto a
single curve.

\section{Square lattice}
\begin{figure}
\includegraphics[clip,width=0.45\textwidth]{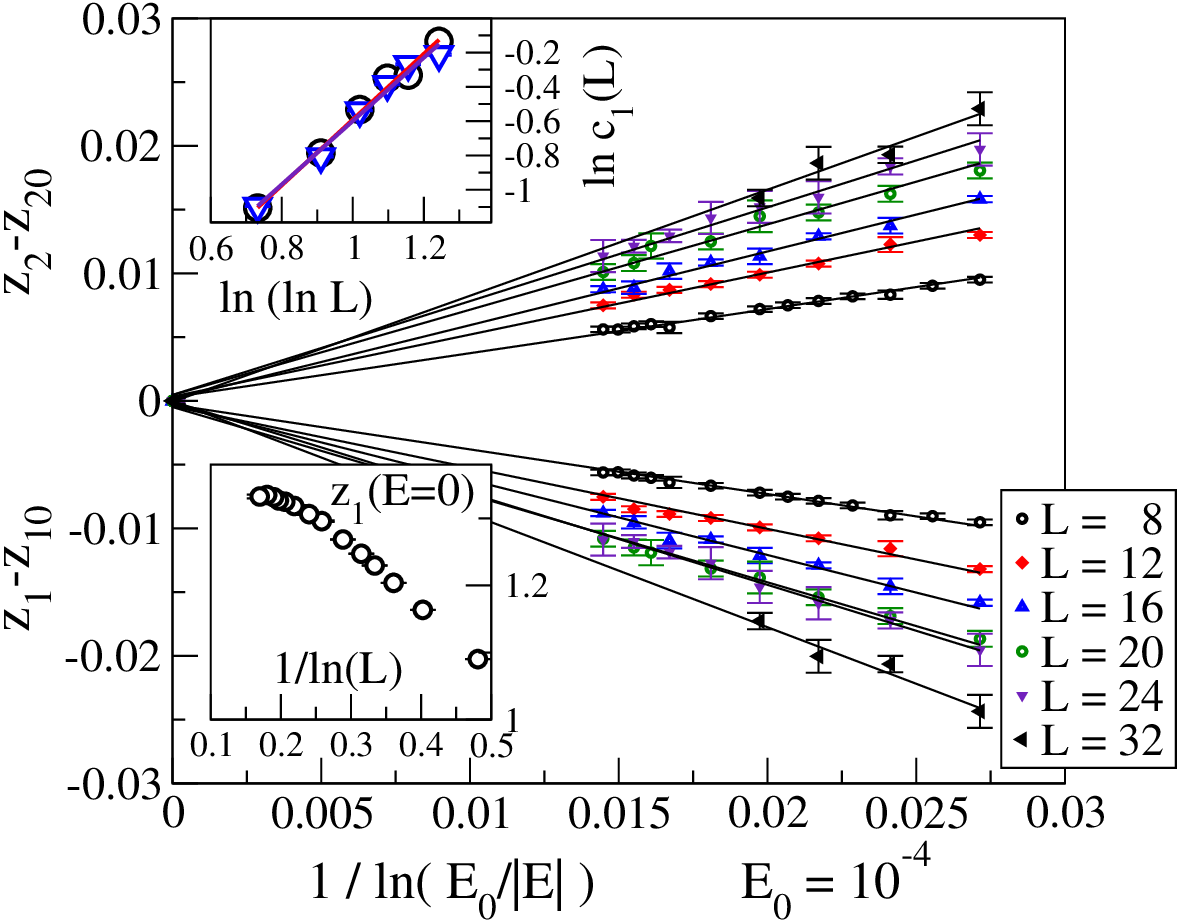}
\caption{(Color online) 
The energy dependence of the two smallest Lyapunov exponents $z_1(E)-z_1(E=0)$
and $z_2-z_2(E=0)$ calculated for the square lattice. The applied random-flux
strength is $f=1.0h/e$. Owing to the small value of $E_0$, we restricted
the energy interval to $|E|\le 10^{-20}$. Then,  both $z_1$ and $z_2$ are
linear functions of $\chi$. The first inset shows the $L$-dependence of
the coefficient $c_1$ of the Taylor expansion (\ref{taylor}). The exponents
$\alpha_1$ obtained are $1.92\pm 0.24$ for $z_1$ and $1.85\pm 0.25$ for $z_2$. 
The plot of $z_1(E=0,L)$ vs. $1/\ln L$ demonstrates the finite size
effects which are much stronger than for the brick layer (lower inset).
}
\label{qq0}
\end{figure}

Another possibility to check for logarithmic scaling at chiral quantum
critical points is the numerical analysis of a simple square
lattice.\cite{Cer00,ERS01,Cer01,ER04,MS07} We calculated $z_1(E,L)$ and
$z_2(E,L)$ for $L$ even and   
Dirichlet boundary conditions in the transverse direction. We found that the
scaling analysis is more difficult than for the bricklayer. First, the finite
size effects are more pronounced (see lower inset in Fig.~\ref{qq0}).
We can eliminate them, at least partially, by subtracting the value $z(E=0,L)$
from $z(E,L)$.\footnote{This does not eliminate the finite size effects
completely, since the value of $z(E=0,L)$ is known only with some uncertainty.}
Second, in the unperturbed model the van Hove singularity, which appears at $E=0$
compared with $E=\pm 1$ for the bricklayer, may spoil the scaling analysis.
More importantly, following the same procedure as for the bricklayer, we
found that the function $X$ given by Eq.~(\ref{XX}) possesses a minimum only
for small  $E_0\sim 10^{-4}$ although the energy band widths are about the
same for both lattices. Since the energy $E$ must be much smaller than
$E_0$, we had to restrict our analysis to energies $|E|\le 10^{-20}$. 
Fortunately, in such a narrow energy interval, we can neglect the
quadratic term in the Taylor expansion (\ref{taylor}). As shown in
Fig.~\ref{qq0}, both $z_1$ and $z_2$ are linear functions of $\chi$ when $L
\le 32$. This enables us to estimate the exponent $\alpha_1$ from the analysis
of the size dependence of the slope, $c_1(L)\propto (\ln L)^{\alpha_1} $. This
analysis is independent on both the choice of $E_0$ and finite size effects,
provided that the latter do not depend on the energy. 

However, the fit turns out to be rather unstable to small changes of the data
ensemble. First, the interval of $\chi$ is very narrow and almost all data
points are accumulated in the right part of this interval. Therefore, the
resulting fit is very sensitive to the exact value of $z_1(E=0)$. Second,
although we calculated our data with high accuracy, Fig.~\ref{qq0} shows that
this is still not sufficient for a perfect determination of the slope. To
check the accuracy of $\alpha_1$, we tested various data ensembles and found
that $\alpha_1$ varies between 1.7 and 2.1. Nevertheless, our data for the
square lattice are compatible with a logarithmic scaling relation.

\section{Conclusions}
We analyzed the scaling behavior of the two smallest Lyapunov exponents $z_1$
and $z_2$ in disordered two-dimensional chiral systems defined on a
bricklayer and on a square lattice. 
We found that both $z_1$ and $z_2$ follow a logarithmic scaling
relation as considered by Sittler and Hinrichsen\cite{SH02} with a correlation  
length proposed by Fabrizio and Castelliani.\cite{FC00} According to
Ref.~\onlinecite{SH02}, the physical origin of logarithmic scaling is
associated with multifractality and local scaling invariance. To the best of
our knowledge, the results presented above are the first numerical
confirmation of the scaling relation (\ref{scal-1}), 
in which the scaling variable is given by the ratio of the 
logarithm of the system width $L$ to the logarithm of the correlation length
$\xi$, instead of the ratio $L/\xi$ as applied usually. This scaling is
accompanied by the logarithmic energy dependence of the correlation length
$\ln\xi(E)\propto \sqrt{\ln (E_0/|E|)}$ valid for $|E|\ll E_0$. Our results also
solve the contradiction between previous numerical work, which apparently did
not reach the true scaling regime, and the Harris criterion. 

Two methods of the scaling analysis were used. Both confirm that the
logarithmic scaling is observable only for very small values of the energy
close to the chiral quantum critical point at $E=0$. In order to resolve the
Lyapunov exponents for energies down to $|E|=10^{-34}$, the implementation of
quadruple precision in the numerical algorithms was necessary.
This probably explains why logarithmic scaling was not observed in
previous numerical work.\cite{Cer00,ERS01,Cer01,ER04,MS07,SM08a} 

The question arises whether the same logarithmic scaling analysis can be
performed also for the two-terminal conductance. At present this seems not
possible with our available computing power. In our previous work,\cite{SM08a}
we found $g(E,L)=g_0\ln(\tilde{E}^{\star}(L)/|E|)$ for 
$|E|>E^{\star}=\tilde{E}^{\star}/\textrm{const.}$, but did not observe any
energy and system size dependence of the ensemble averaged conductance 
$g_c\simeq 4/\pi\,e^2/h$ as long as the energy remains smaller than a certain
value $E^{\star}\propto L^{-2}$. This size dependent energy interval coincides
with the recently observed depression in the density of states.\cite{Sch09}
The observation of a tiny logarithmic energy dependence of the conductance, if
present at all, would require a far more accurate numerical determination of
the ensemble averaged mean conductance.   

As shown analytically in Ref.~\onlinecite{FC00}, the logarithmic energy
dependence of the correlation length near $E=0$ is accompanied by a divergence
of the density of states 
$\rho(E)\propto E^{-1}\exp[-(4A\ln E_0/E)^{1/2}]$. Such a relation is, however,
not found in recent numerical work on a unitary chiral lattice model, where the
density of states decreases to zero when $E\to 0$.\cite{Sch09} 
Also, a different divergency exponent of the density of states
$\rho(E) \propto E^{-1}\exp[- 1/2(c|\ln E/E_0|)^{2/3}]$ was derived analytically
for the chiral orthogonal model.\cite{MDH02,MRF03} This difference shows also
up in the energy dependence of the correlation length.
It would be very interesting to see, whether this subtle difference can also be
observed in numerical studies on a bricklayer lattice with real random hopping
disorder, which belongs to the chiral orthogonal symmetry class.

Acknowledgments: PM thanks project APVV No. 51-003505 and project VEGA 0633/09
for financial support. 

%\bibliographystyle{apsrev}
%\bibliography{paPers_DB,ludwig,qhe}

\end{document}